# NODES: A PROPOSED SOLUTION TO FERMI'S PARADOX

**JOHN GERTZ** Foundation for Investing in Research on SETI Science and Technology (FIRSST), c/o Zorro Productions, 125 University Avenue, Suite 101, Berkeley, CA, 94710 USA
**email:** jgertz@firsst.org

Within the SETI community, a school of thought holds that ET might prefer to send information physically in so-called "probes," rather than by radio or optical beacon, in effect, a message in a bottle. In this paper, a related solution to Fermi's Paradox (also known as the "Great Silence") is proposed whereby ET civilizations aggregate knowledge into a system of "nodes," interspersed throughout the galaxy. Each node explores and serves the local star systems, detecting through its exploration non-technological life, and enlisting newly emergent technological civilizations, such as ourselves, into its central system. Each node would download information to new member civilizations and upload information from those civilizations, passing it along to its immediately adjacent nodes, such that new information would pass through the entire galaxy at near light speed. The most local node would not be directly detectible by us until it signaled Earth in response to a detection of its artificial electromagnetic (EM) leakage. This is because Earth would not be in the narrow beam of the signal pathways between our local node and its adjacent nodes, nor between the local node and other local civilizations, while the spillover signal strength of far distant nodes would be too weak to be detectable. Thus, the Great Silence of the galaxy is simply the result of the fact that Earth is not currently situated within a communications pathway in an otherwise well interconnected galaxy.

**Keywords:** SETI, ET, Nodes, Probes, Fermi's Paradox

## 1. INTRODUCTION

After nearly 60 years of searching, SETI researchers have routinely come up dry. Many solutions to Fermi's classic question, "Where is everybody?," known as Fermi's Paradox or the Great Silence, have been proposed [1]. Arguments have included the possibility (a) that we are alone, at least in the Milky Way, either because we are the first civilization to emerge, or because our predecessors are now extinct; (b) that ET exists but chooses not to contact us, perhaps because they are frightened, incurious, or adhere to an ethic of not wishing to disturb new civilizations; (c) they are transmitting in a manner that is detectable, but we have not found the signal yet simply because it is a very large haystack in which we seek a very small needle; (d) that we are looking in an incorrect fashion, perhaps because they are transmitting in infrared (IR) while so far we are mostly looking in optical and radio bands; or (e) as a subset of (d), that ET is not broadcasting information by means of EM waves, but rather sending information physically on its equivalent of a hard drive, and we have yet to systematically search our own Solar System for these probes. The current paper proposes a different solution, whereby ET's nearest, but remotely positioned, artificially intelligent (AI) outpost intends to contact Earth as soon as it detects us as technologically capable of receiving its signal.

## 2. PROBES RECONSIDERED

### 2.1 Advantages of Probes over Beacons

This author, among others, has previously argued that great advantages would accrue to ET in sending information physically by way of probes instead of by EM beacons from its home planet [2,3,4,5,6,7]. Probes would require vastly less energy to send the same given amount of information relative to continuously broadcasting beacons. Probes completely solve for the problem of L in Drake's equation, the length of time an ET civilization persists in transmitting, since the fate of the progenitor civilization would not affect the probe once it was launched in the direction of our Solar System. Probes would also probably be easier for us to discover. By broadcasting to us locally, their signal may be stronger relative to far distant beacons. Therefore, we would presumably be able to determine the content of a signal and not just the fact of a signal, while we might not be capable of detecting a message within the carrier wave of a far distant beacon using our current generation of telescopes. A probe's signal would probably be persistent, because Earth would be its only target, as opposed to beacons, that might cycle among hundreds to millions of stars and have a very short dwell time on our Solar System. This limiting factor would be multiplied by our own strategy of slewing among thousands of target stars, the resultant combined effect of the two targeting strategies being an extremely low probability of detection.

### 2.2 Disadvantages of probes

These advantages notwithstanding, there would be large inadequacies in this approach from ET's point of view:

- Each civilization would be required to send its own probe to the Solar System. It would obviously be more efficient were civilizations to join together so as not to duplicate efforts.

- Each ET civilization might also feel the need to send multiple probes if they wish to routinely update Earth on the progress of their civilization (while beacons could continuously update their content).

- Probes could lay dormant within our Solar System for eons while they await a signal from Earth that it had be-

come technologically active. That signal might never come, as Earth's biology might never emerge as technological. In the meantime, probes would pile up uselessly in our Solar System.

• The sending civilization would have little realistic chance of receiving a reply. The probe might have arrived in our Solar System millions or billions of years before humans emerged as technologically capable of detecting them, delaying a response by that amount. During such a protracted period of latency, the probe might even have lost all track of its progenitor civilization as both our star and its star meander in their orbits around the center of the Milky Way.

## 3. NODES

### 3.1 Nodes defined

The author initially distinguishes "nodes" from probes by defining the former as positioned away from our Solar System, while the latter as resident within it. Nodes would have characteristics in common with cell phone towers, public libraries, the Internet, server farms, and galactic explorers.

### 3.2 Nodes as cell phone towers and a Galactic Internet

AI nodes might be placed at set distances from one another, each servicing a galactic volume of something on the order of 100 cubic light years (merely a guess for purposes of illustration). Alternatively, they might be interspersed in accordance with the number of stars or perhaps the number of technological civilizations within their service areas; i.e., more nodes in more heavily populated areas of the galaxy. Nodes within a galactic network would constantly update one another, each node being in contact with, for example, the half dozen nodes nearest it, such that new information would circulate throughout the galaxy at almost the speed of light. Nodes would therefore contain identical or near-identical information sets. Redundancy assures the survival of this library system from supernovae, gamma ray bursts, hostile civilizations, or other hazards.

Each node would serve the local community of civilizations. If there are no local civilizations, nodes would still provide invaluable services by acting as signal relay stations or explorers, and by passing on information about potentially habitable planets available for colonization. Because they would be relatively nearby, the energy requirements to send and receive signals would be much reduced relative to what would be required for civilizations to directly communicate with one another from remote corners of the galaxy. Lasers would seem to make the best sense, as over a local distance of ~100LY a mere 1000 W laser would be able to transmit 1012 bits per second in a very narrow beam specifically targeting Earth (i.e., <.02 arc sec., resulting in a beam of <1AU) [8]. Because they are nearby, upload and download intervals would be short – perhaps on the order of less than a century, rather than potentially tens of thousands of years.

### 3.3 The Nodes as explorers and ambassadors

Nodes would be responsible for locating and recruiting into the system emerging technological civilizations, such as our own. Because they are nearby, they might easily detect emergent civilizations by way of their EM leakage.

• Nodes might act as translators. Having databases filled with possibly thousands or millions of civilizations, nodes might be in a position to format information in a fashion most easily understandable to recently emergent civilizations.

• Long before technological civilization would emerge, nodes might have sent tiny flyby probes to surveil each star within its sphere of influence and responsibility along the lines of what is currently contemplated by the Breakthrough Starshot project [9]. Consequently, the nodes would have long foreknowledge that Earth was biologically active.

### 3.4 Establishing a commercial relationship with a node

The nodes may work on a "give a book/get a book" system. Nodes would issue a "library card" to civilizations that agree to share their own proprietary information. That is, civilizations would not be allowed to download unless and until they were willing to upload. In practice, because a node would have orders of magnitude more and better information than the newly applying civilization, it could afford to start the trade by downloading a small subset of information as a "loss leader."

### 3.5 Nodes as arbitrators of galactic behavior

Nodes might also curate downloads, so that, for example, blueprints for nuclear bomb manufacture would be unavailable or restricted, and neither would information on the whereabouts of extant civilizations. The library would constantly grow as more civilizations joined, and member civilizations routinely updated. Nodes, by being the sole contact point for civilizations, could arbitrate behavior. Aggressive or otherwise misbehaving civilizations could be cut off from further communication. Nodes might serve as traffic controllers, allocating or suggesting planets for colonization among member civilizations.

### 3.6 Letting go of our fondness for fellow squishies

It may be that by joining the node system, we would, in effect, lose any ability to directly contact fellow 'squishies' (i.e., carbon based life forms). We would have basically no alternative, and there would be no point to such an effort in any event after contact is achieved with a node. We may emotionally miss a lack of squishy-to-squishy contact, but the node system would completely obviate the need for it, and would actually be superior. It would be much safer for civilizations to communicate through nodes. After all, there could be hostile and dangerous actors in the galaxy who could intercept Earth's exploratory outreach efforts, such as by directing powerful radio beams at random stars (a technique known as METI, short for "messaging ET intelligence"). In common with probes, the node system would be impervious to Drake's L, the length of time that a civilization broadcasts, be that era ended through extinction, boredom, exasperation, change of philosophical outlook, budgetary constraints, or some other cause. The library would continuously grow larger over time, irrespective of how many contributing civilizations persist, or whether their total number increases or decreases over time.

The node system does not disadvantage Earth in any way relative to direct civilization-to-civilization contact. What do we want from ET? First, we wish to simply know that we are not alone. Second, we want them to tell us what they know. Is their math the same as ours; what do we not know about the laws of physics; what do they look and act like; is their biology based upon carbon and some variant of DNA, and so forth? Finally, we want to know about their origin, history (including whether and why they might have gone extinct), and about their "art" and beliefs. All of this could be gained through information exchange with a node.

### 3.7 Who established the node system?

Even after contact, the origin of the node system might prove an enduring mystery. We need not know the exact origin and history of the Internet or Bitcoin to effectively use either. It is even possible that a single, long-extinct civilization initiated the node system, perhaps by launching just one von Neumann replicator [13] eons ago.

### 3.8 Disadvantages of nodes from Earth's point of view

The proposed node system does disadvantage Earth, relative to within-Solar System probes:

- Upload and download times would be measured in decades and not minutes or hours.

- We would lose the ability to retrieve and physically examine a probe.

- Probes might serve as ET ambassadors, that is, as AI beings (i.e., able to pass the Turing test) who might be able to converse with us in real time.

Nevertheless, beggars cannot be choosers. Because we need contact with a node more than it needs contact with us, it is the node that will set the terms of engagement, and probably operate with regard to its own maximal efficiency rather than ours.

### 3.9 Probes and nodes acting in concert

- The Node Hypothesis does not necessarily negate the Probe Hypothesis. For example, once a flyby mission determines that Earth is life bearing, perhaps a node might send a permanent probe to our Solar System. Such a probe might flag to the node Earth's emergence as a technologically active planet, and act as a local ambassador. It could also act as its own relay station, amplifying and directing Earth's EM precisely toward the local node. For example, the local probe might intercept its favorite episode of *I Love Lucy,* extracting it from the cacophony of all of Earth's concurrent transmissions. It might then translate and reformat the episode in a manner that can be understood by nodes. Finally, it would directly and narrowly beam the episode on to the nearest node, which node would in turn incorporate it into its normal data stream, and send it on to other nodes. In such a fashion, *I Love Lucy* would soon permeate the galaxy in a way in which its omnidirectional leakage from Earth would not directly allow.

- Probes and Nodes might be one and the same. A single AI entity situated, for example, on a local asteroid could perform the functions of a probe in its monitoring of and communications with Earth, and at the same time serve as the local node, being in constant communication with nearby nodes in Earth's galactic neighborhood. In effect, the probe/node would have a transmitter/receiver pointed at Earth, with another (or the same) transmitter/receiver pointed at local nodes. Alternatively, McConnell has suggested that extant civilizations might disseminate information circuitously among themselves through a node system, rather than exclusively through direct one-to-one channels of communication [11].

- Although the *I Love Lucy* radius has been in common parlance among SETI researchers for many years, some who have examined its assumptions closely have concluded that it is more mythic than real. They argue that at interstellar distances, although military and planetary radar may be detectable (although they bear no information other than to indicate their technological origin), TV and radio broadcasts (which bear lots of information) would damp down and cancel each other out, such that they would become incoherent and indistinguishable from broadband noise [12,13,14]. In such event, nodes would have no practical alternative but to preposition probes in those solar systems where biologically indicative planets are known to reside if they wish to know something about the content of leaked transmissions.

## 4. SEARCHING FOR NODES

### 4.1 Nodes might be anywhere

Nodes could be anywhere there is an energy source and material to build out capabilities and self-repair, such as on a planet, asteroid, or moon in another stellar system, in orbit around a brown or white dwarf, or perhaps even in interstellar space – possibly associated with a planet that had been previously ejected from its own solar system.

### 4.2 METI as as search strategy

Proponents of METI might argue that nodes merely await our signaled desire to join the galactic club. They could be right. However, for all of the reasons this author, among others, have previously stated, this is a very bad idea [15,16,17]. Even were this author very confident in the correctness of the current hypothesis, it would be unethical to impose his certainty upon everyone else, imperiling all of Earth because of a personal conviction. Moreover, because we currently have no idea where the nearest node may be located, the entire sky would have to be blindly targeted, making it easy for any hostile galactic actor to intercept the signal.

### 4.3 Think galactically, but search locally

First radio transmissions date from around 1920, however, EM leakage really picked up with the widespread dissemination of over-the-air terrestrial television broadcasts. Therefore, the distance that these early signals have travelled is often whimsically referred to in the SETI literature as the *I Love Lucy* radius. As stated above, ET may not be able to discern our leakage as being anything more than broadband noise, but nonetheless, it might reasonably determine it to be technologically generated noise. SETI may be futile beyond .5x this radius. ET will simply contact us when it first detects Earth's artificial EM radiation. There is no need for us to do anything, except to keep our telescopes aimed at local (i.e., < .5 x *I Love Lucy* radius) targets.

The node's signal will probably persist until the node receives Earth's response. The node would correctly reason that it will take our civilization time to discover its signal, time to decode it, time to decide what to do, time to craft a response, and time for that response to return to the node.

It has never been obvious that we should only search locally. If the strategy has been to establish squishy-life-form to squishy-life-form direct communication, then reasonable solutions to the Drake Equation suggest that the nearest such intelligent and communicating civilization could be quite distant from Earth. However, if one abandons the paradigm that the ET one is searching for is a squishy life form and instead devote the search to AI probes or nodes, then their local proximity is a quite reasonable conjecture.

### 4.4 Search Strategies

The node's signal may be so obvious that we will pick it up almost despite ourselves. A laser pointed persistently at us might show up as a bright new "star" in the sky where none existed before. By spectroscopy, the new point source would be instantly recognizable as narrow band and therefore artificial. It might not be associated with a known star, but rather be free floating in interstellar space, or associated with a brown dwarf that would not normally be on any SETI target list – or, indeed, listed in any current star catalogue. Consequently, the signal could be coming from any direction and not just from a known local stellar target. A node might show up in a simple survey of photographic "plates" by searching for any recent change in the light emanating from local (i.e., <~35-50 LY) stars. More broadly, a 360 degree search might be conducted for a new visual object within the <~35-50LY envelope. Breakthrough Listen is currently engaged in comprehensive radio and optical surveys of local stars, including those within half the *I Love Lucy* radius.

If the nearest node is 100 LY from us, it is may detect Earth's artificial EM soon, and we might therefore hear from it in about another century at the earliest. If the nearest node is about 35-50 LY away, it might already be signaling Earth.

In the meantime, there is probably nothing we can do to detect our local node. Nodes would only communicate with their immediately neighboring nodes and the civilizations within their service area, if any. We are therefore unlikely to be in a communications pathway and would not be able to eavesdrop. Nodes might employ very narrow beam lasers not only because they are efficient carriers of information, but also explicitly to avoid signal spillover by which they might be inadvertently detected. Moreover, they might be specifically positioned such that no biologically active planet is either in their foreground or detectable background. This would then explain why the galaxy is not buzzing with detectable signals, and in itself serve as a solution to Fermi's Paradox. The only hope, then, for a pre-contact detection might be in viewing the node's waste heat in the IR. However, unless the node is very near or very large and hot, its heat signature would likely be beneath the detectability level of the James Webb Space Telescope or other current IR detection systems.

## 5. CONCLUSIONS

If the Node Hypothesis is correct, it solves Fermi's Paradox. We are not alone. ET exists, and the galaxy is awash in its signal traffic. We are blind to those signals simply because they are very narrow beam and only pass between nodes and their immediately adjacent nodes, or between nodes and the civilizations within their immediate service areas. Earth is simply not in those locally few and narrow pathways. Signals broadcast by nodes would not be any more powerful than necessary (e.g., 1000 W laser), and over large distances lasers would be obscured by gas and dust clouds. Consequently, we would not be able to detect traffic among nodes nodes that are in other regions of the galaxy.